\documentclass[final]{elsarticle}

\makeatletter
\def\ps@pprintTitle{%
  \let\@oddhead\@empty
  \let\@evenhead\@empty
  \def\@oddfoot{Published in Neurocomputing (https://doi.org/10.1016/j.neucom.2017.08.072)\hfil}
  \let\@evenfoot\@oddfoot
}
\makeatother

\usepackage{lineno,hyperref}

\journal{Neurocomputing}

\usepackage{amssymb}
\usepackage{graphicx}
\usepackage{amsmath}
\usepackage[ruled,noline,linesnumbered]{algorithm2e}
\usepackage[caption=false,font=footnotesize]{subfig}

\newcommand{\bbbr}{\mathbb{R}}

\begin{document}

\begin{frontmatter}

\title{Android Malware Detection with Unbiased Confidence Guarantees}

\author{Harris Papadopoulos\corref{correspondingauthor}}
\ead{h.papadopoulos@frederick.ac.cy}
\cortext[correspondingauthor]{Corresponding author}

\author{Nestoras Georgiou}

\author{Charalambos Eliades}

\author{Andreas Konstantinidis}

\address{Department of Computer Science and Engineering,\\
Frederick University, Cyprus}




\begin{abstract}
The impressive growth of smartphone devices in combination with the rising ubiquity of using mobile platforms for sensitive applications such as Internet banking, have triggered a rapid increase in mobile malware. In recent literature, many studies examine Machine Learning techniques, as the most promising approach for mobile malware detection, without however quantifying the uncertainty involved in their detections.  In this paper, we address this problem by proposing a machine learning dynamic analysis approach that provides provably valid confidence guarantees in each malware detection. Moreover the particular guarantees hold for both the malicious and benign classes independently and are unaffected by any bias in the data. The proposed approach is based on a novel machine learning framework, called Conformal Prediction, combined with a random forests classifier. We examine its performance on a large-scale dataset collected by installing 1866 malicious and 4816 benign applications on a real android device. We make this collection of dynamic analysis data available to the research community. The obtained experimental results demonstrate the empirical validity, usefulness and unbiased nature of the outputs produced by the proposed approach.
\end{abstract}


\end{frontmatter}


\section{Introduction}

The evolution of ubiquitous smartphone devices has given rise to great opportunities with respect to the development 
of applications and services spanning 
from simple messaging and calling applications to more sensitive financial transactions and Internet banking services. 
As a result, a great deal of sensitive information, such as access passwords and credit card numbers, are stored on 
smartphone devices, which has made them a very attractive target to cybercriminals. More specifically, a significant 
increase of malware attacks was observed in the past few years, aiming at stealing private information and sending it 
to unauthorized third-parties.

Mobile malware are malicious software used to gather information and/or gain access to mobile computer devices such as smartphones or tablets. In particular, they are packaged and redistributed with third-party applications to inject malicious content into a smartphone and therefore expose the device's security. While the first one appeared in 2004 targeting the Nokia Symbian OS~\cite{cabir}, in the fourth quarter of 2015 G DATA security experts reported discovering 8,240 new malware applications on average per day and a total of 2.3 million new malware samples in 2015, in just the Android OS~\cite{gdatamalwarereport}. When malware compromises a smartphone, it can illegally watch and impersonate its user, participate in dangerous botnet activities without the user's consent and capture user's personal data.

Mobile malware detection techniques can be classified into two major categories: static analysis~\cite{christ:staticmalware,shabt:staticmalware} and dynamic analysis~\cite{egele:surveydynamic,rieck:learningmalware,mosk:detectbehavior}. The former aim at detecting suspicious patterns by inspecting the source code or binaries of applications. However, malware developers bypass static analysis by employing various obfuscation techniques and therefore limiting their ability to detect polymorphic malware, which change form in each instance of the malware~\cite{moser:limitsstatic}. Dynamic analysis techniques on the other hand, involve running the application and analyzing its execution for suspicious behavior, such as system calls, network access as well as file and memory modifications. The main drawback of these techniques is that it is difficult to determine when and under what conditions the malware malicious code will be executed.

Both static and dynamic analysis techniques are typically implemented following two main approaches: signature-based approaches, which identify known malware based on unique signatures~\cite{griffin:signaturemalware}, and heuristic based approaches, which identify malicious malware behaviour based on rules developed by experts or by employing machine learning techniques~\cite{shabt:staticmalware,mosk:detectbehavior,jacob:behaviorsurvey,menahem:ensemble}. Even though signature-based techniques have been successfully adopted by antivirus companies for malware detection in desktop applications, this is not a preferred solution in the case of mobile devices due to their limited available resources in terms of power and memory. Additionally, signature-based techniques cannot detect zero-day malware (not yet identified) or polymorphic malware, something that is not an issue for heuristic based techniques. On the other hand, unlike signature-based techniques, heuristic based techniques are prone to false positive detections (i.e. wrongly identifying an application as malware).

Most recent research studies focus on extending the idea of heuristic-based approaches by employing machine learning techniques. For example, Sahs and Khan~\cite{Sahs:malware} use a one-class Support Vector Machine to detect malicious applications based on features extracted from Android Application Packages (APKs) of benign applications only. Demertzis and Iliadis~\cite{Demertzis2016} propose a hybrid method that combines Extreme Learning Machines with Evolving Spiking Neural Networks using features extracted from the behaviour of applications when executed on an emulated Android environment. 
In~\cite{Demertzis2015} the same authors propose an extension to the Android Run Time Virtual Machine architecture that analyses the Java classes of applications using the Biogeography-Based Optimizer (BBO) heuristic algorithm for training a Multilayer Perceptron to classify applications as malicious or benign. Abah et. al.~\cite{abah:malware} present a detection system that uses a \emph{k}-Nearest Neighbour classifier to detect malicious applications based on features extracted during execution. In a different kind of study Allix et. al.~\cite{allix:empiricalmalware} study the gap between \emph{in-the-lab} and \emph{in-the-wild} performance of malware detectors. The authors propose a static analysis approach and evaluate it on different settings showing that there is a huge performance drop when malware detectors are tested \emph{in-the-wild}. To the best of our knowledge, none of the machine learning based methods proposed in the literature provides any reliable indication on the likelihood of its detections being correct. The provision of such an indication however, would be of great value for the decision of the user on whether to remove an application or not, depending on the risk he/she is willing to take. 

A recent study by our group \cite{nestoras:malware} examined the utilization of a novel framework, called \emph{Conformal Prediction} (CP) \cite{vovk:alrw}, for quantifying the uncertainty involved in machine learning Android Malware detection. Specifically, CP enabled the provision of provably valid confidence guarantees for each individual prediction without assuming anything more than that the data is exchangeable, a somewhat weaker assumption than the universally accepted in machine learning i.i.d.\ (independent and identically distributed data) assumption. In effect CP transforms the single predictions provided by conventional machine learning techniques into prediction sets (or regions) with a guaranteed error rate, which is at most one minus a pre-specified confidence level. However, the guarantees provided by the standard CP framework hold over all instances together and there is no assurance that they will hold on different categories of instances \cite{papa:StrokeLCCP}. As a result, the particular guarantees will hold over both malicious and benign predictions together, but no guarantee can be provided on malicious detections alone. This is an important issue especially since the inherent class imbalance of malware detection data can lead to a huge bias towards the benign class. Additionally, our previous study was based on dynamic analysis data collected in a controlled emulated environment, which is the typical scenario for studies performed on the detection of Malware.

The aim of this study is twofold: (i) To provide stronger guarantees that hold on malicious and benign instances separately and thus are not biased towards one of the two classes; (ii) Evaluate the performance of the proposed technique on a large-scale realistic dataset.

The first aim is achieved with the use of a modification of the CP framework, called \emph{Label-conditional Mondrian CP} (LCMCP), which guarantees that the required confidence level will be satisfied within each class; i.e. the frequency of errors will be lower than or equal to the required one for the instances of each class separately. In fact we can even require a different confidence level for each class. As the original CP and LCMCP versions of the framework are too computationally demanding for a mobile device application, we follow the inductive version of the CP framework, called \emph{Inductive Conformal Prediction}~\cite{papa:icpnn}, with the Label-conditional Mondrian modification. We combine Label-conditional Mondrian Inductive Conformal Prediction (LCMICP) with a random forests (RF) classifier, which is one of the most popular machine learning techniques. In our experiments we examine the performance of the proposed RF-LCMICP approach and demonstrate its within class validity and its superiority over the conventional RF classifier on which it is based. It should be noted that the proposed LCMICP approach is general in the sense that it can be used for extending any state of the art malware detection technique in order to provide provably valid confidence guarantees. In fact, it can be utilized for providing the same guarantees for any machine learning task involving class imbalance and computational efficiency considerations (such as limited available computation resources or large volumes of data). Furthermore, since the only assumption made by the CP framework is that the data is exchangeable, the validity of the guarantees provided by the proposed approach is not affected by other typical machine learning issues such as outliers and overfitting.

The second aim is accomplished by generating a dataset with state measurements of a real Android device (LG E400) during simulation of random interaction with malicious and benign applications. The dataset consists of state measurements for 6682 applications selected randomly from a large collection of Android applications including a large variety of malware types. The recorded data are made available to aid reproducibility of our results and further research.

The rest of the paper starts with an overview of the Conformal Prediction framework and its Mondrian and inductive counterparts in Section~\ref{sec:cp}. The next section (Section~\ref{sec:rf-lcicp}) details the proposed approach. Section~\ref{sec:data} describes the collection of the data used for evaluating our approach, while Section~\ref{sec:res} presents our experiments and the obtained results. Finally, Section~\ref{sec:conc} gives our conclusions and directions for future work.

\section{Conformal Prediction}\label{sec:cp}

\subsection{The Conformal Prediction Framework}\label{ssec:cpf}

The Conformal Prediction (CP) framework extends conventional machine learning algorithms into techniques that produce 
reliable confidence measures with each of their predictions. The typical classification task consists of a training set 
$\{(x_1, y_1), \dots, (x_l, y_l)\}$ of instances $x_i\in \bbbr^d$ together with their associated classifications 
$y_i \in \{Y_1, \dots, Y_c\}$ and a new unclassified instance $x_{l+1}$. The aim of 
Conformal Prediction is not only to find the most likely classification for the unclassified instance, but to also 
state something about its confidence in each possible classification.

CP does this by assigning each possible classification $Y_j, j = 1, \dots, c$ to $x_{l+1}$ in turn and extending the 
training set with it, generating the set
\begin{equation}
\label{eq:extset}
  \{(x_1, y_1), \dots, (x_l, y_l), (x_{l+1}, Y_j)\}.
\end{equation} 
It then measures how strange, or non-conforming, each pair $(x_i, y_i)$ in (\ref{eq:extset}) is for the rest of the 
examples in the same set. This is done with a \emph{non-conformity measure} which is based on a conventional machine 
learning algorithm, called the \emph{underlying algorithm} of the CP. This measure assigns a numerical 
score $\alpha_{i}^{(Y_j)}$ 
to each pair $(x_i, y_i)$ indicating how much it disagrees with all other pairs in (\ref{eq:extset}). In effect it 
measures the degree of disagreement between the prediction of the underlying algorithm for $x_i$ after being trained 
on (\ref{eq:extset}) with its actual label $y_i$; in the case of $x_{l+1}$, $y_{l+1}$ is assumed to be $Y_j$.

To convert the non-conformity score $\alpha_{l+1}^{(Y_j)}$ of $(x_{l+1}, Y_j)$ into something informative, CP 
compares it with all the other non-conformity scores $\alpha_{i}^{(Y_j)}, i = 1, \dots, l$. This comparison is performed 
with the function
\begin{equation}
\label{eq:pvalue}
  p((x_1, y_1), \dots, (x_l, y_l), (x_{l+1}, Y_j)) = \frac{|\{i = 1, \dots, l : \alpha_{i}^{(Y_j)} \geq \alpha_{l+1}^{(Y_j)}\}|+1}{l+1}.
\end{equation}
The output of this function, which lies between $\frac{1}{l+1}$ and 1, is called the p-value of $Y_j$, also denoted 
as $p(Y_j)$, as this is the only unknown part of~(\ref{eq:extset}). If the data are independent and identically 
distributed (i.i.d.), 
the output $p(y_{l+1})$ for the true classification of $x_{l+1}$ has the property that $\forall \delta\in [0, 1]$ and 
for all probability distributions $P$ on $Z$,
\begin{equation}
\label{eq:validity}
  P^{l+1}\{((x_1,y_1), \dots, (x_{l+1},y_{l+1})):p(y_{l+1}) \leq \delta\}\leq \delta;
\end{equation}
for a proof see \cite{nouretdinov:iid}. Therefore all classifications with a p-value under some very low threshold, 
say $0.05$, are highly unlikely to be correct as such sets will only be generated at most $5\%$ of the time by any 
i.i.d.\ process.

Based on the property (\ref{eq:validity}), given a \emph{significance level} $\delta$, or confidence level $1 - \delta$, 
a CP calculates the p-value of all possible classifications $Y_j$ and outputs the prediction set
\begin{equation}
\label{eq:predregion}
	\{ Y_j : p(Y_j) > \delta \},
\end{equation}
which has at most $\delta$ chance of not containing the true classification of the new unclassified example.
In the case where a single prediction is desired, called \emph{forced prediction}, instead of a prediction set, 
CP predicts the 
classification with the largest p-value, which is the most likely classification, together with a confidence and 
a credibility measure for its prediction. The confidence measure is calculated as one minus the second largest 
p-value, i.e. the significance level at which all but one classifications would have been excluded. This gives an 
indication of how likely the predicted classification is compared to all other classifications. The credibility 
measure on the other hand, is the p-value of the predicted classification. A very low credibility measure indicates 
that the particular instance seems very strange for all possible classifications. In the particular task of 
malware detection low credibility would indicate that an application behaves differently from all known malicious and 
benign applications. This would signify that the application may contain a new type of malware.

\subsection{Label-conditional Mondrian Conformal Prediction}

Conformal Prediction guarantees that the prediction sets it produces will make an 
error with a probability at most as high as the preset significance level (one minus 
the confidence level) over all examples. However, it does not guarantee that this will 
hold on different categories of examples, i.e. it does not provide the same guarantee 
separately for easy and hard examples. Therefore it is possible for prediction sets 
to have a lower than the significance level frequency of errors on easy examples and a 
higher than the significance level frequency of errors on hard examples. Overall one 
will compensate for the other resulting in the correct frequency of errors. This poses a problem 
in the case of class imbalance, which is the case in malware detection as most applications 
are benign, especially since the malicious (minority) class is the most important.

A modification of the original Conformal Prediction framework, called 
\emph{Mondrian Conformal Prediction} (MCP) can guarantee that the resulting prediction sets 
will be valid within categories. Specifically MCP is given a division of the examples into 
categories in the form of a measurable function that assigns a category $\kappa_i$ to each 
example $z_i$. This function is called a \emph{Mondrian taxonomy}. MCP then calculates the 
p-values of the examples in each category separately thus ensuring that validity will hold 
within each category.

A special case of MCP is the Label-conditional MCP (LCMCP), in which the category of each example 
is determined by its label/classification. In this case the p-value of each possible 
classification $Y_j$ of the test example $x_{l+1}$ is calculated with the function
\begin{equation}
\label{eq:pvaluelccp}
  p(Y_j) = \frac{|\{i = 1, \dots, l : y_i = Y_j \,\,\&\,\, \alpha^{Y_j}_i \geq \alpha^{Y_j}_{l+1}\}|+1}{|\{i = 1, \dots, l : y_i = Y_j\}|+1},
\end{equation}
instead of (\ref{eq:pvalue}). Now the prediction set (\ref{eq:predregion}) will make an 
error with a probability at most $\delta$ regardless of the true classification of the example.
Note that in this case we can also use a different significance level for each 
classification and produce the prediction set
\begin{equation}
\label{eq:predregion2}
	\Bigl\{ Y_j : p(Y_j) > \delta_j \Bigr\},
\end{equation}
where $\delta_1, \dots, \delta_c$ are the significance levels corresponding to each classification 
respectively. Forced prediction is calculated in the same way described in Subsection \ref{ssec:cpf} but 
using the p-values produced by (\ref{eq:pvaluelccp}).

\subsection{Label-conditional Mondrian Inductive Conformal Prediction}\label{subsec:icp}

The transductive nature of the original CP framework, including LCMCP, means that all computations have to start from 
scratch for every new test instance. Obviously this is too computationally demanding for a mobile phone application. 
For this reason the proposed approach follows the inductive version of the framework, 
called Inductive Conformal Prediction (ICP), which 
only performs one training phase to generate a general rule with which it can then classify new examples with 
minimal processing.

Specifically, ICP divides the training set (of size~$l$) into the \emph{proper training set} with $m < l$ examples 
and the \emph{calibration set} with $q := l-m$ examples. It then uses the proper training set for training the 
underlying algorithm (only once) and the examples in the calibration set for calculating the p-value of each possible 
classification of the new test example. In effect, after training the underlying algorithm on the proper training set 
the non-conformity scores $\alpha_{m+1}, \dots, \alpha_{m+q}$ of the calibration set examples are calculated. Then 
to calculate the p-value of each possible classification $Y_j \in \{Y_1, \dots, Y_c\}$ of a new test example $x_{l+1}$, 
ICP only needs to calculate the non-conformity score of the pair $(x_{l+1}, Y_j)$ using the already trained underlying 
algorithm and compare it to the non-conformity scores of the calibration set examples with the function
\begin{equation}\label{eq:pvalueicp}
   p(Y_j) = \frac{|\{i = m+1, \dots, m+q : \alpha_i \geq \alpha^{Y_j}_{l+1}\}|+1}{q+1}.
\end{equation}

In the case of the Label-conditional Mondrian
ICP (LCMICP) followed here, the nonconformity score of the pair $(x_{l+1}, Y_j)$ is compared to the nonconformity scores of the 
instances in the calibration set with classification $Y_j$ instead of all instances. In particular, this comparison is 
performed with the function
\begin{equation}\label{eq:pvaluelcicp}
   p(Y_j) = \frac{|\{i = m+1, \dots, m+q : y_i = Y_j \,\,\&\,\, \alpha_i \geq \alpha^{Y_j}_{l+1}\}|+1}{|\{i = m+1, \dots, m+q : y_i = Y_j\}|}.
\end{equation}
Notice that the steps that need to be repeated for each test example have almost negligible computational requirements.
Again the prediction sets and forced prediction outputs are calculated in the same way as LCMCP.

\section{Proposed Approach}\label{sec:rf-lcicp}

In this study, LCMICP was combined with Random Forests (RF) as underlying algorithm. The RF classifier consisted of 100 
decision trees trained on different bootstrap samples of the proper training set with a randomly selected subset of 
attributes; the number of attributes selected for each tree was equal to the square root of the number of total attributes. 
RF was implemented using the TreeBagger class of the Matlab statistics and machine learning toolbox \cite{matlabstats}. 

When given an instance $x_i$ to classify, a trained decision tree $h_t$ produces the posterior probability $\hat P_t(Y_j|x_i)$ 
for each classification $Y_j$, defined as the number of training instances with classification $Y_j$ that lead to the 
same node as $x_i$ divided by the total number of training instances that lead to that node. The RF classifier averages 
the posterior probabilities produced by all its decision trees to produce the posterior probability
\begin{equation}
    \hat P(Y_j|x_i) = \frac{1}{T}\sum_{t=1}^{T} \hat P_t(Y_j|x_i),
\end{equation}
where $T$ is the number of decision trees in the RF ensemble. In this work 
the classification task is binary, therefore $Y_j \in \{0, 1\}$ and two probabilistic values are produced: 
$\hat P(0|x_i)$ and $\hat P(1|x_i)$.

The nonconformity measure used for the proposed RF-LCMICP is
\begin{subequations}\label{eq:nm}
\begin{align}
	\alpha^{Y_j}_i &= 1 - \hat P(y_i|x_{m+i}), \mspace{15 mu} i = 1, \dots, q, \label{eq:nm1} \\ 
	\alpha^{Y_j}_{l+1} &= 1 - \hat P(Y_j|x_{l+1}), \label{eq:nm2}
\end{align}
\end{subequations}
where $\hat P(y_i|x_i)$ is the RF posterior probability for the true classification of $x_i$ and 
$\hat P(Y_j|x_{l+1})$ is the RF posterior probability for the assumed class $Y_j \in \{0, 1\}$ of $x_{l+1}$.

The complete process followed by the proposed approach is detailed in Algorithm~\ref{al:rflcmicp}. Lines~1 to 5 correspond to the training phase that needs to be performed only once. This phase trains the RF on the proper training set (line~1) and calculates the nonconformity score of each calibration example $x_{m+i}$, $i = 1,\dots,q$, by inputing it to the trained RF to obtain the probabilistic outputs $\hat P(0|x_{m+i})$ and $\hat P(1|x_{m+i})$ (line~3) and using them in (\ref{eq:nm1}) to calculate $\alpha_{m+i}$ (line~4). The testing phase, lines~6 to 11, is the only part that needs to be repeated for every new instance. This phase obtains the probabilistic outputs of the trained RF for the new instance (line~6) and then for each possible class, it calculates the corresponding nonconformity score with (\ref{eq:nm2}) in line~8 and uses it together with the nonconformity scores of the calibration examples in (\ref{eq:pvaluelcicp}) to calculate the p-value of the new instance belonging to that class (line~9). Finally, in line~11, it outputs the prediction set (\ref{eq:predregion}).

\begin{algorithm}[t]
\KwIn{proper training set $\{(x_1, y_1), \dots, (x_m, y_m)\}$, \\
\hspace{28.5pt} calibration set $\{(x_{m+1}, y_{m+1}), \dots, (x_{m+q}, y_{m+q})\}$, \\
\hspace{28.5pt} test example $x_{l+1}$, ensemble size $T$, significance level $\delta$}
$H = {h_1, \dots, h_T} \leftarrow$ train the RF classifier on $\{(x_1, y_1), \dots, (x_m, y_m)\}$\;

\For{$i = {1}$ \KwTo $q$}
{
   $\{\hat P(0|x_{m+i}), \hat P(1|x_{m+i})\} \leftarrow H(x_{m+i})$\;
   $\alpha_{m+i} \leftarrow 1 - \hat P(y_{m+i}|x_{m+i})$\;
}
$\{\hat P(0|x_{l+1}), \hat P(1|x_{l+1})\} \leftarrow H(x_{l+1})$\;

\For{$Y_j = 0$ \KwTo $1$}
{
   $\alpha^{Y_j}_{l+1} \leftarrow 1 - \hat P(Y_j|x_{l+1})$\;
   
   $p(j) = \frac{|\{i = m+1, \dots, m+q : y_i = Y_j \,\,\&\,\, \alpha_i \geq \alpha^{Y_j}_{l+1}\}|+1}{|\{i = m+1, \dots, m+q : y_i = Y_j\}|}$\;
}
\KwOut\\
\Indp
\Indp
Prediction set $R \leftarrow \Bigl\{ Y_j : p(Y_j) > \delta \Bigr\}$.
\caption{Binary RF-LCMICP\label{al:rflcmicp}}
\end{algorithm}

\section{Data Collection}\label{sec:data}

For evaluating the proposed approach a dataset was generated by installing Android application files (\emph{.apk}) on a LG E400 Android device and recording its states while running them and simulating user interaction. Specifically, the examined .apk files were obtained from the AndroZoo collection of android applications\footnote{Available online at: https://androidzoo.uni.lu} of the University of Luxembourg, which has analyzed each application by utilizing several anti-malware tools in order to classify them as malicious or benign. Our dataset consists of state recordings for 6682 applications from the aforementioned collection, of which 1866 are malicious and 4816 are benign. The data recorded for each application include Binder, Battery, Memory, CPU, Network, and Permission information similarly to \cite{amos:dynamic}. Table~\ref{tab:features} presents all the recorded features. Here it is important to note that the Battery related data was not considered in our experiments, due to the fact that the smartphone device was continually charging. Additionally all Diff related features (Binder: TotalNodesDiff, TotalRefDiff, TotalDeathDiff, TotalTransactionDiff, TotalTransactionCompleteDiff; and Network: TotalTXPacketsDiff, TotalTXBytesDiff, TotalRXPacketsDiff, TotalRXBytesDiff) were not used in our experiments as they can be derived from other features (see the feature combinations in Subsection~\ref{ssec:exppre}).

The data collection was performed by selecting the .apk files in a random order, snuffling in this way the malicious and benign applications being installed. For each .apk file, the state of the smartphone device was recorded before and while simulating random user interactions with the application using the Android ``\emph{adb-monkey}'' tool. For the simulated user interaction 1200 touch events were performed with an interval of 450ms while the state of the smartphone device was recorded every $5$ seconds. At the end of the interaction the application was terminated and uninstalled. For avoiding interference among the effects caused by applications on the state of the device a one minute delay was applied between removing an application and initiating the process for the next one. Additionally, the device was restarted after recording data for five applications.

The data collected for each application (and .apk file names) are available at: https://github.com/harrisp/malware-data.

\begin{table}[t]
  \centering
  \caption{Recorded features divided into categories.}
  \label{tab:features}
  \begin{tabular}{l@{\hskip 10pt} l} \hline\noalign{\smallskip}
   Category & Features \\ 
  \noalign{\smallskip}\hline\noalign{\smallskip}
   Battery    & IsCharging, Voltage, Temp, Level, LevelDiff \\
   Binder     & Transaction, Reply, Acquire, Release, ActiveNodes, \\
			  & TotalNodes, ActiveRef, TotalRef, ActiveDeath, TotalDeath, \\
              & ActiveTransaction, TotalTransaction, \\ 
              & ActiveTransactionComplete, TotalTransactionComplete, \\
              & TotalNodesDiff, TotalRefDiff, TotalDeathDiff, \\
              & TotalTransactionDiff, TotalTransactionCompleteDiff \\
    \noalign{\smallskip}\hline\noalign{\smallskip}
  CPU         & User, System, Idle, Other \\
    \noalign{\smallskip}\hline\noalign{\smallskip}
  Memory      & Active, Inactive, Mapped, FreePages, AnonPages, FilePages, \\
              & DirtyPages, WritebackPages \\
    \noalign{\smallskip}\hline\noalign{\smallskip}
  Network     & TotalTXPackets, TotalTXBytes, TotalRXPackets, \\
              & TotalRXBytes, TotalTXPacketsDiff, TotalTXBytesDiff, \\
              & TotalRXPacketsDiff, TotalRXBytesDiff \\
    \noalign{\smallskip}\hline\noalign{\smallskip}
  Permissions & TotalPermissions\\
    \noalign{\smallskip}\hline\noalign{\smallskip}
  \end{tabular}
\end{table}

\section{Experimental Results}\label{sec:res}

\subsection{Data Preprocessing and Experimental Setting}\label{ssec:exppre}

In our experiments we combined the data collected for each application into one set of features by calculating each 
feature value in six different ways:
\begin{itemize}
    \item Mean: The mean of all values collected during interaction with the application.
    \item MeanDiff: The difference between Mean and the previous state of the device.
    \item MedianDiff: The difference between the median of all values collected during interaction and previous state 
          of the device.
    \item MinDiff: The difference between the minimum of all values collected during interaction 
          and the previous state of the device.
    \item MaxDiff: The difference between the maximum of all values collected during interaction 
          and the previous state of the device.
    \item Std: The standard deviation of all values collected during interaction.
\end{itemize}
Note that in all cases each application corresponds to one instance in the dataset, unlike the dataset used 
in \cite{amos:dynamic}.

The performance of the proposed RF-LCMICP approach and of its underlying RF technique was examined on each of the 
six feature sets individually with an ensemble of 100 trees. Before each experiment all input features were normalized 
to the range [0,1].

We performed two groups of experiments varying the percentage of malicious applications in the training set to examine 
the effect of class imbalance to the results. In both groups of experiments the test set was comprised of 300 benign and 
300 malicious applications selected randomly from the collected data. The training set in the first group of experiments 
consisted of 4500 benign and 1500 malicious applications while in the second group it consisted of 4500 
benign and 500 malicious applications, therefore malicious applications made up $25\%$ and $10\%$ of the training 
set respectively. In the case of RF-LCMICP, the calibration set was formed by randomly selecting $20\%$ of the training 
examples of each class minus one so that the denominator in equation (\ref{eq:pvaluelcicp}) would be a multiple 
of $100$. Specifically, in the first 
group of experiments the calibration set consisted of 899 benign and 299 malicious applications, while in the second group 
it consisted of 899 benign and 99 malicious applications. It is worth noting that in all our experiments the conventional 
RF technique was trained on the whole training set, unlike the underlying model of the RF-LCMICP, which was trained 
on the smaller set resulting from the removal of the calibration instances. 

All experiments were 
repeated 100 times with different randomly selected instances (without replacement) for both forming the training and 
test sets and dividing the training sets into proper training and calibration sets. The results reported here are the 
average values over the 100 repetitions. This ensured that the obtained results are not dependent on a particular 
allocation of instances.

\subsection{Forced Prediction}\label{subsec:acc}

\begin{table}[t]
  \centering
  \caption{Forced prediction performance of the RF-LCMICP and conventional RF when malicious applications make up 
           $25\%$ of the training set.}
  \label{tab:acc1}
  \begin{tabular}{l r c c c c} \hline\noalign{\smallskip}
            \multicolumn{1}{l}{Technique} & \multicolumn{1}{c}{Feature} & Accuracy & Sensitivity & Specificity & F$_1$-score \\
                                          & \multicolumn{1}{c}{Set}     & (\%)     & (\%)        & (\%)        &             \\
  \noalign{\smallskip}\hline\noalign{\smallskip}
          & Mean       & 73.70 & 73.93 & 73.48 & 0.7374 \\
          & MeanDiff   & 76.79 & 76.94 & 76.63 & 0.7681 \\
RF-       & MedianDiff & 76.94 & 77.30 & 76.58 & 0.7702 \\
LCMICP    & MinDiff    & 74.51 & 74.85 & 74.17 & 0.7458 \\
          & MaxDiff    & 71.12 & 71.03 & 71.20 & 0.7107 \\
          & Std        & 73.77 & 73.95 & 73.59 & 0.7380 \\
  \noalign{\smallskip}\hline\noalign{\smallskip}
         & Mean       & 67.23 & 39.85 & 94.62 & 0.5482 \\
         & MeanDiff   & 69.92 & 45.06 & 94.78 & 0.5993 \\
Convent. & MedianDiff & 70.42 & 46.15 & 94.70 & 0.6090 \\
RF       & MinDiff    & 67.61 & 41.46 & 93.77 & 0.5610 \\
         & MaxDiff    & 62.63 & 29.44 & 95.82 & 0.4403 \\
         & Std        & 66.52 & 37.22 & 95.82 & 0.5260 \\
   \noalign{\smallskip}\hline\noalign{\smallskip}
  \end{tabular}
\end{table}

\begin{table}[t]
  \centering
  \caption{Forced prediction performance of the RF-LCMICP and conventional RF when malicious applications make up 
           $10\%$ of the training set.}
  \label{tab:acc2}
  \begin{tabular}{r r c c c c} \hline\noalign{\smallskip}
            \multicolumn{1}{l}{Technique} & \multicolumn{1}{c}{Feature} & Accuracy & Sensitivity & Specificity & F$_1$-score \\
                                          & \multicolumn{1}{c}{Set}     & (\%)     & (\%)        & (\%)        &             \\
  \noalign{\smallskip}\hline\noalign{\smallskip}
         & Mean       & 71.83 & 72.42 & 71.24 & 0.7197 \\
         & MeanDiff   & 73.73 & 74.23 & 73.22 & 0.7384 \\
RF-      & MedianDiff & 73.95 & 74.70 & 73.20 & 0.7412 \\
LCMICP   & MinDiff    & 72.22 & 72.51 & 71.94 & 0.7228 \\
         & MaxDiff    & 67.23 & 67.67 & 66.79 & 0.6734 \\
         & Std        & 70.90 & 71.61 & 70.18 & 0.7108 \\
  \noalign{\smallskip}\hline\noalign{\smallskip}
         & Mean       & 54.02 & \phantom{0}8.60 & 99.43 & 0.1571 \\
         & MeanDiff   & 56.01 & 12.65           & 99.38 & 0.2228 \\
Convent. & MedianDiff & 56.60 & 13.90           & 99.30 & 0.2421 \\
RF       & MinDiff    & 52.80 & \phantom{0}5.99 & 99.60 & 0.1122 \\
         & MaxDiff    & 52.12 & \phantom{0}4.46 & 99.78 & 0.0850 \\
         & Std        & 54.54 & \phantom{0}9.53 & 99.54 & 0.1729 \\
   \noalign{\smallskip}\hline\noalign{\smallskip}
  \end{tabular}
\end{table}

Our first set of experiments evaluates the proposed technique in terms of forced predictions. That is 
when the RF-LCMICP predicts the most likely classification together with a confidence and credibility measure to that 
classification. Tables~\ref{tab:acc1} and~\ref{tab:acc2} report the accuracy, sensitivity, specificity and 
F$_1$-score of the proposed approach and its underlying technique for each feature set when the malicious application
instances compose $25\%$ and $10\%$ of the training set respectively.

The results in these two tables show that the conventional RF technique is highly biased towards the benign 
class and this bias greatly increases when the proportion of malicious instances is reduced. This is evident from 
the huge difference between its sensitivity and specificity values as well as the very low F$_1$-scores. On the 
contrary, the sensitivity and specificity values of the proposed approach are very close and are not affected 
by the degree of class imbalance. Of course the performance of RF-LCMICP is lower in Table~\ref{tab:acc2}, but this 
is expected since it has less malicious training examples to learn from. 

In terms of the six different feature sets, it seems that MeanDiff and MedianDiff give the best performance. It is 
also evident by comparing the performances when using the Mean and MeanDiff features that taking into account the 
state of the device before running the application is beneficial. 

It is worth to point out that in addition to providing unbiased predictions, the main advantage of the proposed approach 
is that it also accompanies these predictions with confidence and credibility measures, something that is not evident in 
the results reported so far. These measures can help identify the cases where predictions are likely to be wrong and 
cases that seem very different from the training data indicating the possibility of a new malware kind. The validity and 
quality of the outputs produced by RF-LCMICP are the subject of the rest of this Section.

\subsection{Empirical Validity}

\begin{figure*}
	\centering
		\subfloat[MeanDiff with $25\%$ Malicious]{\includegraphics[trim = 0.1cm 0cm 0.2cm 0.2cm, clip, width=6cm]{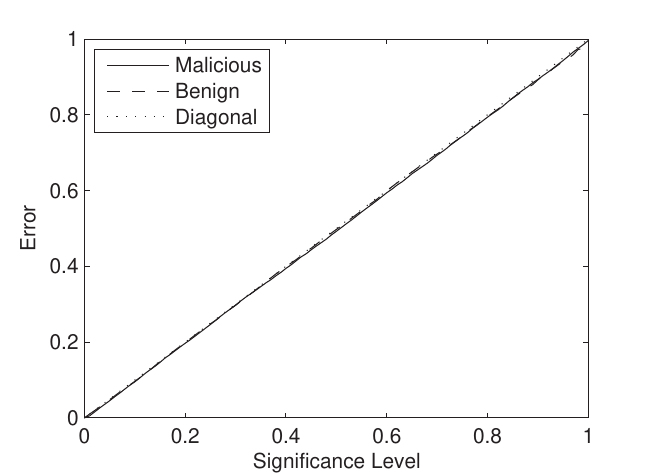}}
		\subfloat[MeanDiff with $10\%$ Malicious]{\includegraphics[trim = 0.1cm 0cm 0.2cm 0.2cm, clip, width=6cm]{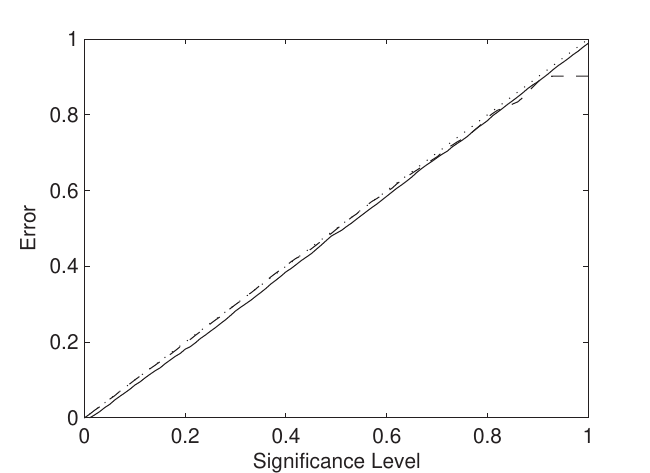}}\\
		\subfloat[MedianDiff with $25\%$ Malicious]{\includegraphics[trim = 0.1cm 0cm 0.2cm 0.2cm, clip, width=6cm]{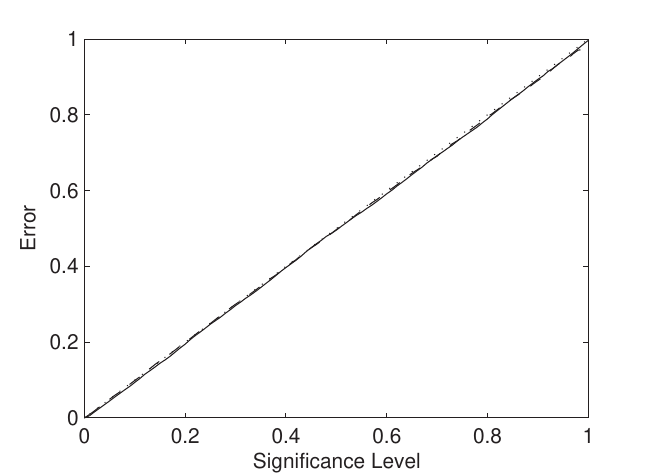}}
		\subfloat[MedianDiff with $10\%$ Malicious]{\includegraphics[trim = 0.1cm 0cm 0.2cm 0.2cm, clip, width=6cm]{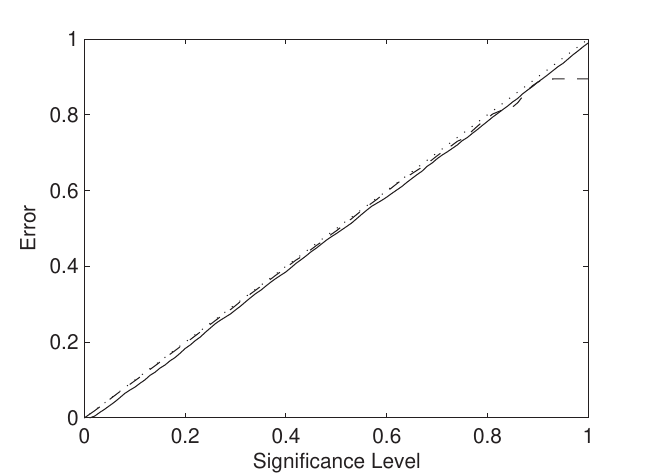}}\\
		\subfloat[MaxDiff with $25\%$ Malicious]{\includegraphics[trim = 0.1cm 0cm 0.2cm 0.2cm, clip, width=6cm]{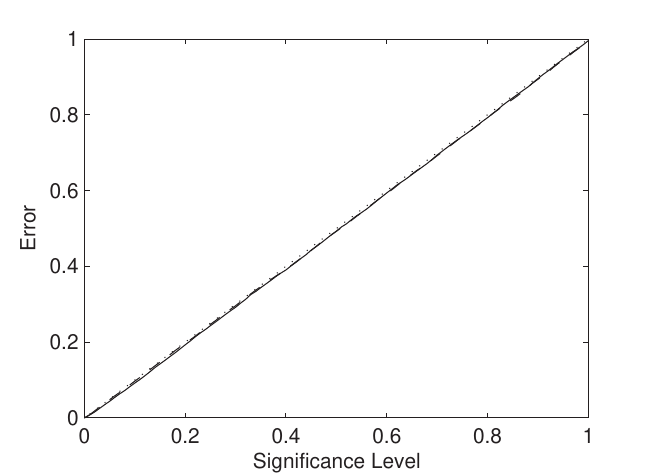}}
		\subfloat[MaxDiff with $10\%$ Malicious]{\includegraphics[trim = 0.1cm 0cm 0.2cm 0.2cm, clip, width=6cm]{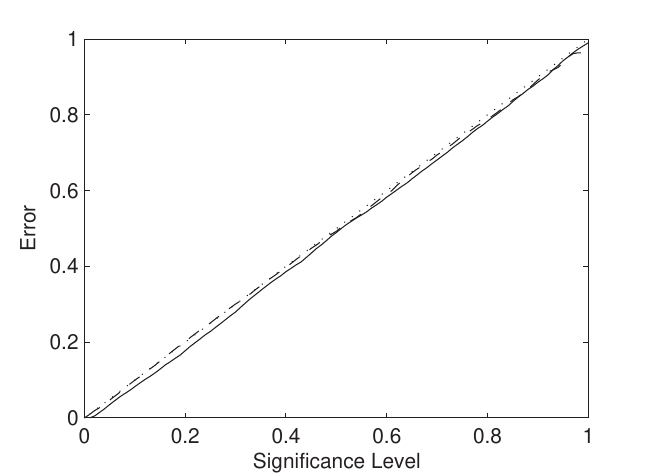}}
\caption{Empirical validity of the RF-LCMICP on the malicious (minority) class and on the 
         benign (majority) class separately. Each pane plots the error percentage of the positive instances 
         with a solid line, the error percentage of the negative instances with a dashed line and the 
         diagonal (exact validity) with a dotted line.}
\label{fig:validityLCICP}
\end{figure*}

\begin{figure*}
	\centering
		\subfloat[MeanDiff with $25\%$ Malicious]{\includegraphics[trim = 0.1cm 0cm 0.2cm 0.2cm, clip, width=6cm]{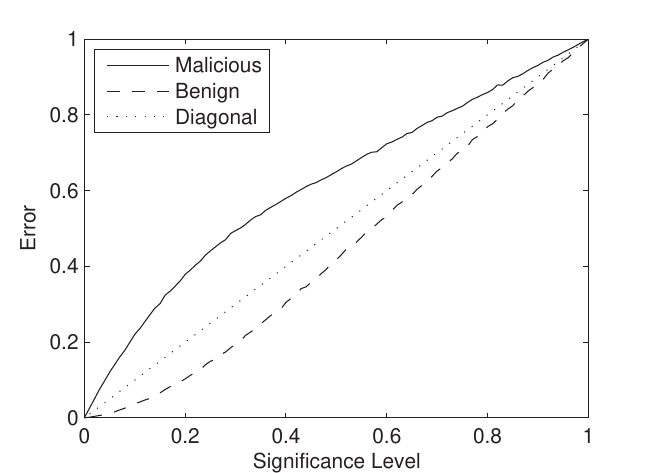}}
		\subfloat[MeanDiff with $10\%$ Malicious]{\includegraphics[trim = 0.1cm 0cm 0.2cm 0.2cm, clip, width=6cm]{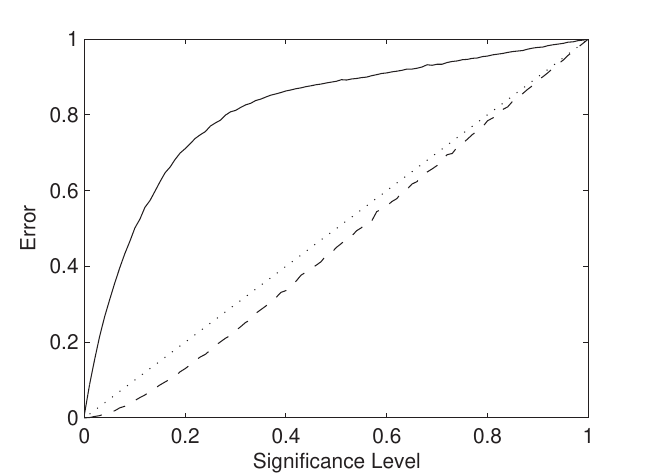}}\\
		\subfloat[MedianDiff with $25\%$ Malicious]{\includegraphics[trim = 0.1cm 0cm 0.2cm 0.2cm, clip, width=6cm]{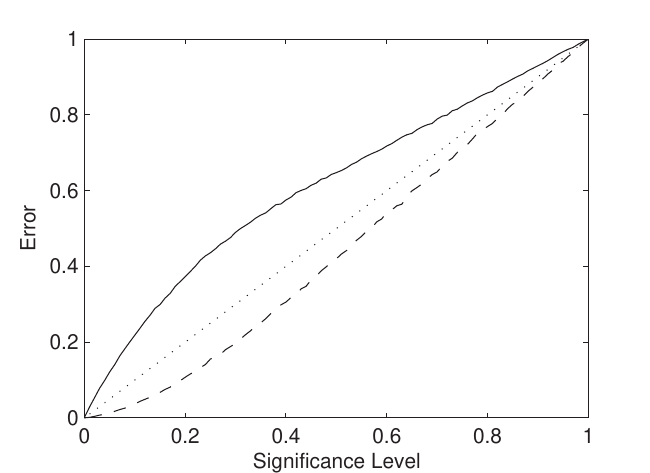}}
		\subfloat[MedianDiff with $10\%$ Malicious]{\includegraphics[trim = 0.1cm 0cm 0.2cm 0.2cm, clip, width=6cm]{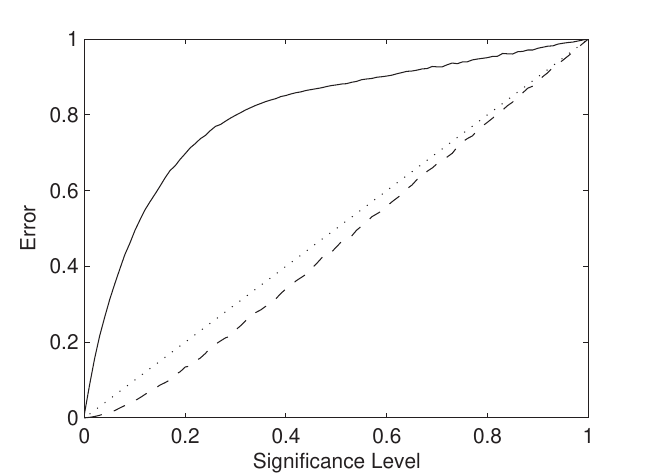}}\\
		\subfloat[MaxDiff with $25\%$ Malicious]{\includegraphics[trim = 0.1cm 0cm 0.2cm 0.2cm, clip, width=6cm]{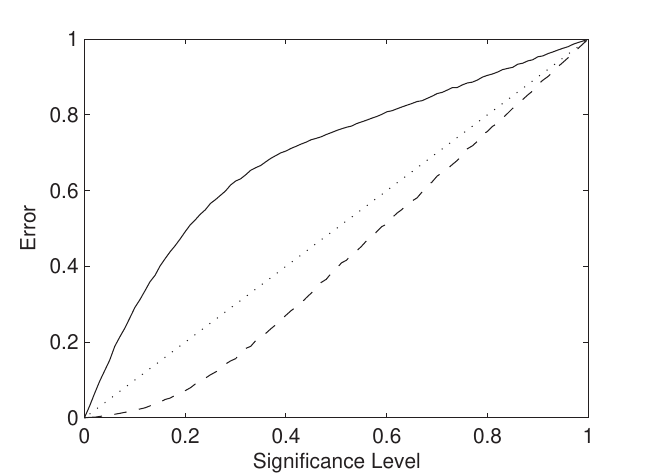}}
		\subfloat[MaxDiff with $10\%$ Malicious]{\includegraphics[trim = 0.1cm 0cm 0.2cm 0.2cm, clip, width=6cm]{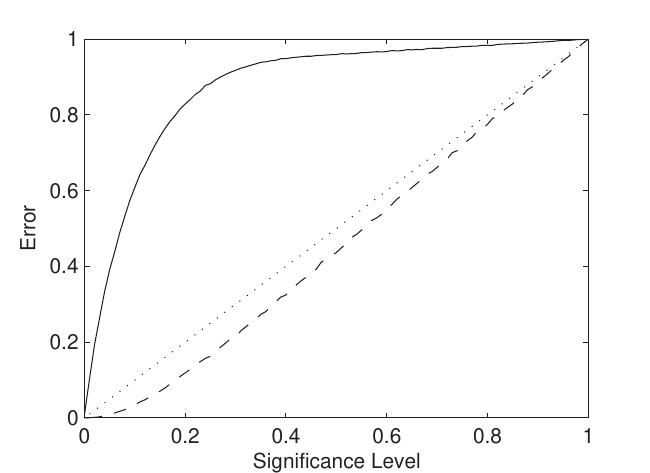}}
\caption{Empirical validity of the conventional RF technique on the malicious (minority) class and on the 
         benign (majority) class separately. Each pane plots the error percentage of the positive instances 
         with a solid line, the error percentage of the negative instances with a dashed line and the 
         diagonal (exact validity) with a dotted line.}
\label{fig:validityOrig}
\end{figure*}

As the provision of unbiased guarantees is a major motivation of this work, this subsection examines the empirical 
validity of the proposed RF-LCMICP and compares it to that of the probabilistic outputs produced by the conventional 
RF technique. The probabilistic outputs of conventional RF were converted to prediction sets by including in each 
prediction set the classification with the highest probability and adding the 
other classification with a probability $(1 - \delta - \hat p(Y_j|x_i))/\hat p(Y_k|x_i)$, where $1-\delta$ is the 
required confidence level, $Y_j$ is the classification with the highest probability and $Y_k$ is the other 
classification. Consequently on average the sum of the probabilities included in the resulting prediction sets 
is $1-\delta$ and thus, given the true conditional probabilities, these prediction sets will contain the true 
classification with a probability $1-\delta$.

Figure~\ref{fig:validityLCICP} plots the error percentages of the prediction sets produced by RF-LCMICP for the 
malicious (minority) instances and the benign (majority) instances. In order to avoid showing extremely similar 
figures for the different feature sets, we selected the two best performing sets (MeanDiff and MedianDiff) and the 
worst performing set (MaxDiff) so as to show that empirical validity is not affected by the quality of the features 
used. The two top plots correspond to the MeadDiff set, the two middle plots correspond to the MedianDiff set and the 
two bottom plots correspond to the MaxDiff set. For each feature set the plots on the left correspond to the training 
sets with $25\%$ malicious instances, while the plots on the right correspond to the training sets with $10\%$ 
malicious instances. In each plot the solid line is the error percentage on malicious instances, the dashed line is 
the error percentage on benign instances and the dotted line represents the diagonal, i.e. where the error rate is 
equal to the significance level $\delta$ (exact validity). In all cases both the solid and the dashed lines are very 
close to the diagonal. This shows that the produced prediction sets make errors on malicious and benign instances 
with the same frequency, which is equal to the required significance levels. Additionally this is not affected by 
the degree of class imbalance in the data. The only difference is a small deviation of the benign error percentages 
from the diagonal at significance levels near $1$ in the more imbalanced cases, but the prediction sets remain valid 
as the errors become less than the corresponding significance levels. Moreover this only happens at significance 
levels above $0.9$ corresponding to confidence levels below $1\%$, which are of no practical interest. 
Overall this figure confirms empirically that the RF-LCMICP guarantees within class validity regardless of the 
quality of the data used or the degree of imbalance.

Figure~\ref{fig:validityOrig} plots the error percentages 
for the probabilistic outputs of the conventional RF technique in the same manner and for the same sets of features. 
Unlike the plots of Figure~\ref{fig:validityLCICP}, the error percentages on the malicious (minority) instances are 
higher than the corresponding significance levels while the errors on the benign (majority) instances are lower. 
This shows once again the bias of conventional RF towards the benign (majority) class. 
Additionally the difference increases in the case of the MaxDiff features and in the right plots when the class 
imbalance is higher. It is also evident that even when considering all instances together the error percentages 
will be higher than the corresponding significance level since on an equal number of malicious and benign test 
instances the deviation of the solid line from the diagonal is bigger than the opposite deviation of the dashed line.

One can appreciate the important consequences that the bias towards the benign class shown in 
Figure~\ref{fig:validityOrig} can have. It would mean mistakenly considering a malicious application as benign with 
high confidence while the opposite is true, therefore providing false security to the user.

\subsection{Quality of p-values}

\begin{table}[t]
  \centering
  \caption{OU-criterion for the RF-LCMICP when malicious applications make up $25\%$ of the training set.}
  \label{tab:ou1}
  \begin{tabular}{r c c c} \hline\noalign{\smallskip}
     \multicolumn{1}{l}{Feature} & \multicolumn{3}{c}{Observed Unconfidence} \\
     \multicolumn{1}{l}{Set}     & All & Malicious & Benign \\
  \noalign{\smallskip}\hline\noalign{\smallskip}
      Mean       & 0.1788 & 0.1771 & 0.1806 \\
      MeanDiff   & 0.1530 & 0.1504 & 0.1556 \\
      MedianDiff & 0.1521 & 0.1497 & 0.1546 \\
      MinDiff    & 0.1765 & 0.1743 & 0.1788 \\
      MaxDiff    & 0.2260 & 0.2248 & 0.2272 \\
      Std        & 0.1832 & 0.1817 & 0.1846 \\
  \noalign{\smallskip}\hline\noalign{\smallskip}
  \end{tabular}
\end{table}

\begin{table}[t]
  \centering
  \caption{OU-criterion for the RF-LCMICP when malicious applications make up $10\%$ of the training set.}
  \label{tab:ou2}
  \begin{tabular}{r c c c} \hline\noalign{\smallskip}
     \multicolumn{1}{l}{Feature} & \multicolumn{3}{c}{Observed Unconfidence} \\
     \multicolumn{1}{l}{Set}     & All & Malicious & Benign \\
  \noalign{\smallskip}\hline\noalign{\smallskip}
      Mean       & 0.2115 & 0.2060 & 0.2170 \\
      MeanDiff   & 0.1903 & 0.1855 & 0.1952 \\
      MedianDiff & 0.1877 & 0.1825 & 0.1930 \\
      MinDiff    & 0.2089 & 0.2037 & 0.2140 \\
      MaxDiff    & 0.2736 & 0.2685 & 0.2786 \\
      Std        & 0.2261 & 0.2201 & 0.2321 \\
  \noalign{\smallskip}\hline\noalign{\smallskip}
  \end{tabular}
\end{table}

\begin{table*}[t]
  \centering
  \caption{N-criterion (average prediction set size) for the RF-LCMICP when malicious applications make up $25\%$ of the training set.}
  \label{tab:n1}
  \begin{tabular}{l r c c c c} \hline\noalign{\smallskip}
             & \multicolumn{1}{c}{Feature} &  \multicolumn{4}{c}{Confidence Level (1 - $\delta$)} \\ 
             & \multicolumn{1}{c}{Set}     & $95\%$ & $90\%$ & $85\%$ & $80\%$ \\
  \noalign{\smallskip}\hline\noalign{\smallskip}
          & Mean       & 1.5790 & 1.3918 & 1.2528 & 1.1380 \\
          & MeanDiff   & 1.5422 & 1.3391 & 1.1958 & 1.0799 \\
All       & MedianDiff & 1.5345 & 1.3394 & 1.1954 & 1.0770 \\
Instances & MinDiff    & 1.5898 & 1.3846 & 1.2416 & 1.1246 \\
          & MaxDiff    & 1.6966 & 1.5093 & 1.3493 & 1.2132 \\
          & Std        & 1.5905 & 1.4022 & 1.2590 & 1.1406 \\
  \noalign{\smallskip}\hline\noalign{\smallskip}
          & Mean       & 1.5889 & 1.4068 & 1.2638 & 1.1441 \\
          & MeanDiff   & 1.5319 & 1.3294 & 1.1903 & 1.0762 \\
Malicious & MedianDiff & 1.5199 & 1.3270 & 1.1877 & 1.0744 \\
Instances & MinDiff    & 1.6105 & 1.4018 & 1.2511 & 1.1285 \\
          & MaxDiff    & 1.6523 & 1.4661 & 1.3208 & 1.1991 \\
          & Std        & 1.5671 & 1.3832 & 1.2439 & 1.1317 \\
  \noalign{\smallskip}\hline\noalign{\smallskip}
          & Mean       & 1.5691 & 1.3769 & 1.2418 & 1.1320 \\
          & MeanDiff   & 1.5524 & 1.3488 & 1.2012 & 1.0836 \\
Benign    & MedianDiff & 1.5492 & 1.3519 & 1.2032 & 1.0796 \\
Instances & MinDiff    & 1.5692 & 1.3674 & 1.2320 & 1.1207 \\
          & MaxDiff    & 1.7408 & 1.5525 & 1.3778 & 1.2274 \\
          & Std        & 1.6139 & 1.4213 & 1.2741 & 1.1496 \\
  \noalign{\smallskip}\hline\noalign{\smallskip}
  \end{tabular}
\end{table*}

\begin{table*}[t]
  \centering
  \caption{N-criterion (average prediction set size) for the RF-LCMICP when malicious applications make up $10\%$ of the training set.}
  \label{tab:n2}
  \begin{tabular}{l r c c c c} \hline\noalign{\smallskip}
             & \multicolumn{1}{c}{Feature} &  \multicolumn{4}{c}{Confidence Level (1 - $\delta$)} \\ 
             & \multicolumn{1}{c}{Set}     & $95\%$ & $90\%$ & $85\%$ & $80\%$ \\
  \noalign{\smallskip}\hline\noalign{\smallskip}
          & Mean       & 1.6709 & 1.4613 & 1.3181 & 1.1949 \\
          & MeanDiff   & 1.6328 & 1.4215 & 1.2747 & 1.1473 \\
All       & MedianDiff & 1.6293 & 1.4162 & 1.2718 & 1.1471 \\
Instances & MinDiff    & 1.6722 & 1.4665 & 1.3140 & 1.1841 \\
          & MaxDiff    & 1.7609 & 1.5820 & 1.4330 & 1.3021 \\
          & Std        & 1.6999 & 1.4968 & 1.3489 & 1.2200 \\
  \noalign{\smallskip}\hline\noalign{\smallskip}
          & Mean       & 1.6274 & 1.4456 & 1.3118 & 1.1893 \\
          & MeanDiff   & 1.5780 & 1.3968 & 1.2595 & 1.1386 \\
Malicious & MedianDiff & 1.5738 & 1.3837 & 1.2500 & 1.1353 \\
Instances & MinDiff    & 1.6469 & 1.4554 & 1.3089 & 1.1833 \\
          & MaxDiff    & 1.7096 & 1.5339 & 1.3957 & 1.2761 \\
          & Std        & 1.6400 & 1.4615 & 1.3260 & 1.2041 \\
  \noalign{\smallskip}\hline\noalign{\smallskip}
          & Mean       & 1.7145 & 1.4770 & 1.3243 & 1.2005 \\
          & MeanDiff   & 1.6876 & 1.4463 & 1.2899 & 1.1559 \\
Benign    & MedianDiff & 1.6849 & 1.4488 & 1.2936 & 1.1590 \\
Instances & MinDiff    & 1.6974 & 1.4776 & 1.3191 & 1.1850 \\
          & MaxDiff    & 1.8123 & 1.6301 & 1.4702 & 1.3282 \\
          & Std        & 1.7599 & 1.5322 & 1.3717 & 1.2359 \\
  \noalign{\smallskip}\hline\noalign{\smallskip}
  \end{tabular}
\end{table*}

This subsection evaluates the quality of the p-values produced by the proposed RF-LCMICP and therefore the informativeness of its 
outputs. In effect we would like the p-values of the incorrect classes to be as low as possible, which will lead to the 
incorrect class being excluded from the resulting prediction sets as soften as possible and to higher confidence 
measures for the true class in the case of forced predictions. This evaluation is performed following two of the 
probabilistic criteria recently proposed in \cite{vovk:cpcriteria}. The first is a $\delta$\emph{-free} and \emph{observed} 
criterion, meaning 
that it does not depend on the significance level $\delta$ and that it takes into account the true classes of instances. 
This is the \emph{observed unconfidence criterion} (OU-criterion), which is the mean of the incorrect class p-values 
regardless of the classification assigned by the technique. The second criterion is a $\delta$\emph{-dependent} 
and \emph{prior} 
criterion, meaning that it depends on the significance level $\delta$ and that it does not take the true classes of 
instances into consideration. This is the average size of the resulting prediction sets for each significance level, 
defined as the \emph{N-criterion} in \cite{vovk:cpcriteria}. Since the error rates of prediction sets are 
guaranteed this criterion 
evaluates the size that these sets need to have in order to satisfy the required significance level.

Table~\ref{tab:ou1} reports the values of the OU-criterion for all instances (first column) as well as for the malicious and 
benign instances separately (second and third column) with each of the six sets of features when $25\%$ of the training 
set instances were malicious. Table~\ref{tab:ou2} reports the same values for the training set with higher class imbalance - 
with only $10\%$ of instances being malicious. In both tables the values for the malicious and benign instances are 
very close, indicating that there is no bias towards one or another. As expected performance worsens when having less 
malicious instances to learn from, but the values remain balanced. Comparing the values of the different sets of 
features one can see that the MeanDiff and MedianDiff perform better than the others, which is consistent with their 
performance in Tables~\ref{tab:acc1} and~\ref{tab:acc2}. The observed unconfidence values of these two feature sets 
correspond to an average confidence close to $85\%$ to the correct class and this includes the cases where the 
forced prediction is wrong. This means that quite a few of the malicious applications can be detected with a confidence 
higher than $85\%$.

Tables~\ref{tab:n1} and~\ref{tab:n2} report the N-criterion values for the confidence levels of 
$95\%$, $90\%$, $85\%$ and $80\%$ (corresponding to $\delta$ set to $0.05$, $0.1$, $0.15$ and $0.2$) 
for the training sets made up of $25\%$ and $10\%$ malicious instances respectively. These tables are divided into three 
parts: the top part reports the average prediction set sizes obtained for all instances together, while the middle and 
bottom parts report the average prediction set sizes for the malicious and benign instances separately. Each part 
contains the average sizes of the prediction sets obtained with each of the six sets of features. As expected, lowering 
the required level of confidence results in smaller prediction sets as it allows a higher error rate. The same 
observations with those from Tables~\ref{tab:ou1} and~\ref{tab:ou2} also apply here: (i) there is no bias towards 
benign or malicious 
instances and this is not affected by the feature set used or the degree of class imbalance, (ii) the best performing 
feature sets are MeanDiff and MedianDiff. Considering the complexity of the particular task, reflected by the 
values reported in Tables~\ref{tab:acc1} and~\ref{tab:acc2}, the average prediction set sizes obtained with the 
two best performing feature sets are arguably a good result. With $95\%$ confidence (allowing only $5\%$ errors) we 
can be sure for close to half of the applications about whether they are malicious or benign. Lowering the required 
confidence level results in smaller sizes, with the $80\%$ confidence level, which is still higher than the obtained 
accuracy, resulting in an average prediction set size close to $1$.

\section{Conclusions}\label{sec:conc}

We propose a machine learning approach for Android malware detection that unlike conventional machine learning based malware detection techniques produces confidence measures in each of its predictions, which are guaranteed to be valid for  malicious and benign instances separately. The proposed approach is based on the combination of the Mondrian and Inductive versions of the Conformal Prediction framework, which produces provably valid confidence measures that have a clear probabilistic interpretation without assuming anything more than i.i.d.\ data.

The proposed approach was evaluated on a large dataset of state recordings obtained by installing a variety of malicious and benign applications on a real android device, thus making the data as close to a realistic setting as possible. Additionally, six different ways of combining the state recordings taken before and during interaction with each application were examined in an effort to identify the best one. The collected dynamic analysis data are made available at: https://github.com/harrisp/malware-data.

Our experimental results on the collected data show that the proposed approach gives predictions that are unbiased towards malicious or benign applications regardless of the degree of class imbalance in the training data, unlike its underlying technique. Furthermore, they demonstrate empirically the within class validity of LCMICP, which means that one can effectively control the frequency of errors on malicious and benign instances separately. On the contrary, the probabilistic predictions produced by the conventional RF technique were shown to be highly biased and therefore misleading. Finally, the confidence measures and prediction sets produced by the proposed RF-LCMICP approach were shown to be quite informative with an average confidence close to $85\%$ in the correct class and an average prediction set size close to $1$ at the $80\%$ confidence level.

Given the complexity of the particular task considering the large variety of both malicious and benign applications and the uncontrolled environment of a real Android device, it is naturally expected that some applications are difficult to detect. As shown by our experimental results, the proposed approach can provide an unbiased estimate on the degree to which a given detection can be expected to be correct, enabling a user to take informed decisions on whether to remove an application or not, depending on the risk he/she is willing to take. Specifically, in a practical usage scenario the user could set a threshold on his/her acceptable confidence for an application not being malicious (i.e. a threshold on the p-value of the malicious class) and if the chance of an application being malicious exceeds that threshold, the user could be notified and presented with the classification and confidence values for it, enabling him/her to decide on removing it or not.

Our immediate future plans include the examination of feature selection and of more complex ways of combining the collected features. Furthermore, the collection and examination of additional features, possibly including static analysis data, is another future goal.

\end{document}